\documentclass[peerreview]{IEEEtran}
\renewcommand{\baselinestretch}{1.5}
\usepackage[dvips]{graphicx}
\usepackage{amsmath,amssymb}
\newcommand{\defeq}{\stackrel{\triangle}{=}}

\newtheorem{definition}{Definition}

\newtheorem{lemma}{Lemma}
\newtheorem{theorem}{Theorem}
\newtheorem{corollary}{Corollary}

\newcommand{\qed}{\fbox{}}

\newif\ifhoge
\hogetrue

\begin{document}

\title{Non-Adaptive Group Testing \\ based on Sparse Pooling Graphs}
\author{Tadashi Wadayama 
\thanks{T. Wadayama is with 
  Nagoya Institute of Technology, Nagoya City,  Aichi, 466-8555, JAPAN.
    (e-mail:wadayama@nitech.ac.jp).
  A part of this work is to be presented at International Symposium on Information Theory, 2013.
  The initial version of this work has been  included in e-preprint server arXiv since Jan. 2013 
  (identificator:{\em 	 arXiv:1301.7519}). 
} }
\maketitle

\begin{abstract}
In this paper, an information theoretic analysis on non-adaptive group testing schemes based on sparse pooling graphs is presented.
The binary status of the objects to be tested are modeled by i.i.d. Bernoulli random variables with probability $p$.
An $(l, r, n)$-regular pooling graph is a bipartite graph with 
left node degree $l$ and right node degree $r$, where $n$ is the number of left nodes.
Two scenarios are considered: a noiseless setting and a noisy one.
The main contributions of this paper are  direct part theorems that  give conditions for the existence of an estimator achieving 
arbitrary small estimation error probability.
The direct part theorems are proved by averaging  an upper bound on estimation error probability of the typical set estimator over 
an $(l,r, n)$-regular pooling graph ensemble.  Numerical results  indicate sharp threshold behaviors in the asymptotic regime.
\end{abstract}

\section{Introduction}

The paper by Dorfman \cite{Dorfman} introduced the idea of {\em group testing} and also 
presented a simple analysis, which indicates advantages of the idea. 
His main motivation was to devise an {\em economical way} to detect infected persons within a population by using blood tests.
It is assumed that the outcome of a blood test determines if the blood used in the test contains certain target viruses (or bacteria).

Of course, blood tests for every person in the population would clearly distinguish the infected individuals 
from those who are not infected.
Dorfman's idea for reducing the number of tests is the following. 
We first divide the population into several disjoint groups and then mix the blood of individuals in each group to form {\em pools}.
The  test process then consists of two-stages. 
In the first stage,  the pools containing infected blood are determined by blood test of each pool.
In the second stage, all the individuals in those groups with positive results are tested.
Numerical examples show that the number of tests can be reduced  without loss of detection capability \cite{Dorfman}.

Dorfman's idea triggered  the emergence of subsequent theoretical works on group testing
and  a variety of practical applications,   such as the screening of DNA clone libraries
and the detection of faulty  machines parts \cite{Du1} \cite{Du2}. 
In addition, recent advances in the theory  of compressed sensing \cite{robust} \cite{LPdecoding}
have stimulated research into the  theoretical aspects of group testing.

The group testing scheme due to Dorfman can be classified as {\em adaptive group testing},
in which the latter part of test design depends on the results of earlier tests. 
There is also  {\em non-adaptive group testing}, in which 
the test design  is completely determined before conducting any tests.
Intuitively, adaptive group testing seems advantageous over non-adaptive group testing
because it requires fewer tests.
However,  there are also advantages to the non-adaptive group testing, since in this design,
all the tests can be executed in parallel.
Note that  adaptive group testing requires 
sequential tests and thus prevents parallel testing.

In order to develop a non-adaptive group testing scheme with good detection performance,  
pool design is crucial.
In the field of combinatorial group testing, 
a pooling matrix that defines the set of pools to be tested  is constructed by using combinatorial design and combinatorics. 
The {\em deterministic construction} of a  $K$-disjunct matrix 
is one of the central themes of  combinatorial group testing \cite{Du1} \cite{Du2}. 

Pooling matrices can also be obtained by {\em random construction}; that is,
the $(0,1)$-elements of a pooling matrix are determined probabilistically.
Several reconstruction algorithms 
have been proposed for such probabilistically constructed pooling matrices.
For example,  Sejdinovic and Johnson \cite{Sejdinovic}, Kanamori et al. \cite{Kanamori} recently 
proposed reconstruction algorithms based on belief propagation.
Malioutov and Malyutov \cite{Malioutov},  Chan et al. \cite{Chan} studied reconstruction algorithms 
based on linear programming (LP).

Clarifying the scaling behavior of the number of required tests for correct reconstruction has become
one of the most important topics in this field.
Berger and Levenshtein \cite{Berger} studied a two-stage group testing scheme and unveiled 
the scaling law for the number of required tests based on information theoretic arguments. 
M\'ezard and Toninelli \cite{Mezard} provided a novel analysis of two-stage schemes based on theoretical techniques 
from statistical mechanics. Recently, Atia and Saligrama \cite{Atia} presented  an information theoretic analysis of non-adaptive 
group testing with and without noise. They presented a direct part theorem that gives a condition for the existence of an estimator achieving 
arbitrary small estimation error probability and a converse part theorem that gives a condition for the non-existence of good estimators.
The arguments in their proof of these theorems are based on the proof of the channel coding theorems for multiple access channels, 
and they can be applied to 
both noiseless and noisy observations. For example, in the noiseless case, it was shown that a $K$-sparse instance of $n$-objects can be 
perfectly recovered from the test results if the number of tests is asymptotically $O(K \log n)$.

The main motivation of this work is to provide an information-theoretic 
analysis of non-adaptive group testing based on sparse pooling graphs.
In this paper, we assume that the status (0 or 1) of an object  is modeled by a Bernoulli random variable with 
probability $p$. In other words, we consider the scenario in which the sparsity parameter $K$ scales as $K \simeq p n$ asymptotically.
In most conventional information theoretic analyses, such as \cite{Atia}, $K$ is assumed to be independent of $n$. Such an assumption is 
reasonable in order to clarify the dependency of the required number of tests on the sparsity parameter and the number of objects.
Although our assumption is different from the conventional one, it is also natural from an information theoretic point of view and is suitable 
for observing  sharp threshold behaviors in the asymptotic regime.

Another new aspect is that the analysis is carried out under the assumption of an $(l,r,n)$-regular 
pooling graph ensemble, which is a bipartite graph ensemble with 
left node degree $l$ and right node degree $r$, where $n$ is the number of left nodes.
This model is suitable for handling a very sparse pooling matrix and is amenable to ensemble analysis.
We will present both direct and converse theorems that predict the asymptotic behavior 
of a group testing scheme with an $(l,r,n)$-pooling graph. 
These asymptotic conditions are parameterized by $p$, $l$, and $r$.
Therefore, for a given pair $(l,r)$, we can determine the region for $p$ in which we can achieve arbitrarily accurate estimation. 
Our analysis was inspired by the analysis of Gallager and others \cite{Gal63}  \cite{LS02} \cite{Hu} of
low-density parity-check (LDPC) codes.

The outline of this paper is organized as follows.
Section \ref{prelimi} provides definition of two group testing systems which are called the noiseless system and the noisy system.
Section \ref{lbound} presents  lower bounds on estimation error probability. These bounds are proved by using Fano's inequality.
Section \ref{directpart} discusses the direct part theorems. 
Section \ref{generalize} describes a generalization of the converse and direct part theorems for a general class of 
a {\em sparse observation system}.

\section{Preliminaries}
\label{prelimi}
In this section, we introduce the two scenarios for group testing that will be discussed in this paper.
The first one is the {\em noiseless system}, where test results can be seen as a function of an input vector.
The second one is the {\em noisy system}, where the test results are disturbed by the addition of noise.

\subsection{Problem setting for the noiseless system}

The random variable $X \defeq (X_1,\ldots, X_n)$ represents the status of $n$-objects.
We assume that $X_i (i \in [1,n])$ is an i.i.d. Bernoulli random variable with 
the probability distribution $Pr(X_i=0)=1-p,  Pr(X_i=1)=p (0\le p \le 1)$. The notation 
$[a,b]$ represents the set of consecutive integers from $a$ to $b$.  With some slight abuse of notation,
the notation $[a,b]$ is also used for representing closed interval over $\Bbb R$ when there is no fear of 
confusion.
A realization of $X$ is denoted by $x \defeq (x_1,\ldots, x_n)$. The test function
$OR(z_1,\ldots,  z_r): \{0,1\}^r \rightarrow \{0,1\}$ is the logical OR (disjunctive) function with 
$r$-arguments ($r$ is a positive integer) defined by
\begin{equation}
OR(z_1,\ldots,  z_r) \defeq
\left\{
\begin{array}{ll}
0, & z_1=z_2=\cdots = z_r = 0 \\
1, & \mbox{otherwise}. 
\end{array}
\right.
\end{equation}
The results of pooling tests which is abbreviated as {\em test results}  are represented by 
$Y \defeq (Y_1, \ldots, Y_m)$. A realization of $Y$ is denoted by 
$y \defeq  (y_1, \ldots, y_m)$.

Let $G \defeq (V_L, V_R ,E)$ be a bipartite graph, called a {\em pooling graph},  with the following properties.
The $n$-nodes in $V_L$  are called {\em left nodes} and the other 
$m$-nodes in $V_R$ are called {\em right nodes}. The set $E$ represents the set of edges.
For convenience, we assume that the left nodes are labeled 
from $1$ to $n$. The left node with label $i \in [1,n]$ 
corresponds to $X_i$;  for simplicity, we will refer to it as left node $i$.
In a similar manner, the right nodes are labeled from $1$ to $m$.
In this paper, $G$ is assumed to 
be an $(l, r,n)$-regular bipartite graph, which means that any left and right nodes
have degrees $l$ and $r$, respectively, 
and that the number of the left nodes is $n$.

\begin{figure}[htbp]
\ifhoge
\begin{center}
\includegraphics[scale=0.35]{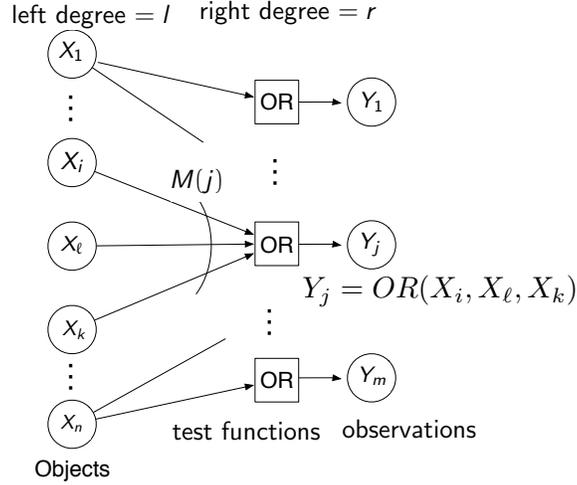}
\end{center}
\fi
\caption{Noiseless system}
\label{fg1}
\end{figure}

For the right node $j \in [1,m]$, the neighbor set of the node $j$ is defined by
$
M(j) \defeq \{i \in [1,n] \mid (i,j) \in E \ \}.
$
We are now ready to describe the relationship between $X$ and $Y$.
For a given pooling graph $G$,  $Y_j (j \in [1,m])$ are related to $X_i (i \in [1,n])$ by
$
Y_j = OR(X_i)_{i \in M(j)}.
$
The notation $(X_i)_{i \in M(j)}$ represents $(X_{j_1}, \ldots, X_{j_r} )$ when $M(j) = \{X_{j_1}, \ldots, X_{j_r}\}$.
Namely, a pooling graph $G$ defines a function from $X$ to $Y$.
We will denote this relationship as $Y=F_G(X)$ for short.
Figure \ref{fg1} illustrates the system configuration of the noiseless system.

The goal of an examiner to infer, as correctly as possible,  the realization of a hidden random variable $X$
from the test observation $y$.
Assume that the examiner uses an estimator (i.e., estimation function) $\Phi: \{0,1\}^m \rightarrow \{0,1\}^n$
for the inference. The estimator  gives an estimate of $x$, 
$
\hat x = \Phi(y),
$
from the test observation $y$. The estimator $\Phi$ should be chosen so that  the {\em estimation error probability} 
\begin{equation}
P_e \defeq Pr\left( \Phi(F_G(X)) \ne X \right)
\end{equation}
is as small as possible.

\subsection{Problem setting for the noisy system}

The setting for the noisy system is almost the same as the setting for the noiseless system, which was
described in the previous subsection. The crucial difference between the two is the assumption of observation noises in the noisy system.
In this case, the examiner observes a realization of the random variable $Z$, defined by
\begin{equation}
Z = Y+ E = F_G(X) + E,
\end{equation}
where $E \defeq (E_1,\ldots, E_m)$ represents the observation noise. 
We assume that
$E_i (i \in [1,m])$ is also an i.i.d.\ Bernoulli random variable with 
the probability distribution $Pr(E_i=0)=1-q,  Pr(E_i=1)=q (0\le q \le 1)$.


\section{Converse Part Analysis}
\label{lbound}

In this section, lower bounds on estimation error probability for the noiseless and noisy systems 
will be shown.  The key to the proofs is Fano's inequality, which ties the estimation error probability to 
the conditional entropy.

\subsection{Lower bound for noiseless system}

Fano's inequality is an inequality that relates the conditional entropy  to the estimation error probability 
and it has often been used as the main tool  in the proof of of the converse part of a channel coding theorem \cite{Cover}.
This inequality plays also a crucial role in the following analysis, in which it clarifies the limit of accurate estimation for 
the noiseless and noisy systems.
\begin{lemma}[Fano's inequality]
Assume that random variables $A, B$ are given.
The cardinalities of the domains (alphabets) of $A$ and $B$ are assumed to be finite.
For any estimator $\phi$ for estimating the hidden value of $A$ from 
the observation of $B$,  the  inequality 
\begin{equation}
1 + Pr(A \ne \phi(B)) \log_2 |{\cal A} |  \ge H(A|B)
\end{equation}
holds. The domain of $A$ is denoted by ${\cal A}$. \hfill \qed
\end{lemma}

We use Fano's inequality for deriving a lower bound on the error 
probability of an estimation for the noiseless system.
Note that this lower bound does not depend on the choice  of  pooling graph 
and an estimator. The proof of the theorem resembles the proof of the upper bound on 
code rate for LDPC codes \cite{Gal63} \cite{Burshtein2}. Similar argument can be found in \cite{Atia}, \cite{Chan} as well.
\begin{theorem}[Lower bound on estimation probability: Noiseless system] \label{th-noiseless}
Assume the noiseless system.
For any pair of an $(l,r,n)$-pooling graph and an estimator, 
the error probability $P_e$ is bounded from below by
\begin{equation} \label{asymbound}
h(p) - \frac{l}{r} h((1-p)^r) - \frac{1}{n} \le P_e.
\end{equation}
\end{theorem}
(Proof) 
For any estimator having the error probability $P_e$, we have
\begin{eqnarray} \nonumber
H(X) &=& I(X;Y) + H(X|Y) \\  \label{duetofano}
&\le& I(X;Y) + 1 + P_e \log_2 |{\cal X}| \\  \label{equation1}
&=& H(Y) - H(Y|X) + 1 + P_e n \\  \label{equation2}
&=& H(Y) + 1 + P_e n.
\end{eqnarray}
The inequality (\ref{duetofano}) is due to Fano's inequality. Equation (\ref{equation1}) holds 
since ${\cal X} = \{0,1\}^n$. 
Note that, in the noiseless system,  the random variable $Y$ is  
a function of $X$, namely $Y = F_G(X)$ and that it implies $H(Y|X)=0$.
The last equality (\ref{equation2}) is a consequence of $H(Y|X)=0$.

Since we have assumed that $X=(X_1,\ldots, X_n)$ is an $n$-tuple of  i.i.d.\   Bernoulli random variables,
the entropy of $X$ is given by $H(X) = n h(p)$, where $h(p)$ is the binary entropy function defined by
$h(p) \defeq -p \log_2 p -(1-p) \log_2(1-p)$.
We thus have
\begin{equation}
n h(p) \le H(Y) + 1 + P_e n.
\end{equation}

Next, we need to evaluate $H(Y) = H(Y_1, \ldots, Y_m)$. 
It should be noted that the random variables $Y_1, Y_2, \ldots, Y_n$ are binary random variables, and
they are correlated in general. 
A simple upper bound on $H(Y)$ can be obtained as 
\begin{eqnarray} \nonumber
H(Y_1,Y_2,\ldots, Y_m) &=& H(Y_1) + H(Y_2|Y_1)  + \cdots + H(Y_m|Y_{m-1},\ldots, Y_1) \\  \label{Ybound}
&\le& H(Y_1) + H(Y_2) + H(Y_3) + \cdots + H(Y_m). 
\end{eqnarray}
This is simply due to the chain rule and a property of the conditional probabilities
(i.e., conditioning reduces entropy \cite{Cover}).

From our assumptions that $Y_j = OR(X_i)_{i \in M(j)} (j \in [1,m])$ and that $|M(j)| = r(j \in [1,m])$, 
we have $H(Y_j) = h((1-p)^r)$  because $Pr[Y_j = 0] = (1-p)^r$.
Combining the inequality (\ref{Ybound}) and $H(Y_j) = h((1-p)^r)$,  we obtain an inequality 
\begin{equation} \label{ineqep}
n h(p) \le m h((1-p)^r) + 1 + P_e n.
\end{equation}

From inequality (\ref{ineqep}), we immediately obtain a lower bound on 
the error probability $P_e$ as 
\begin{equation}
h(p) - \frac{l}{r} h((1-p)^r) - \frac{1}{n}  \le P_e,
\end{equation}
where the relationship $m/n = l/r$ is used. \hfill\qed

\begin{figure}[htbp]
\begin{center}
\includegraphics[scale=1.0]{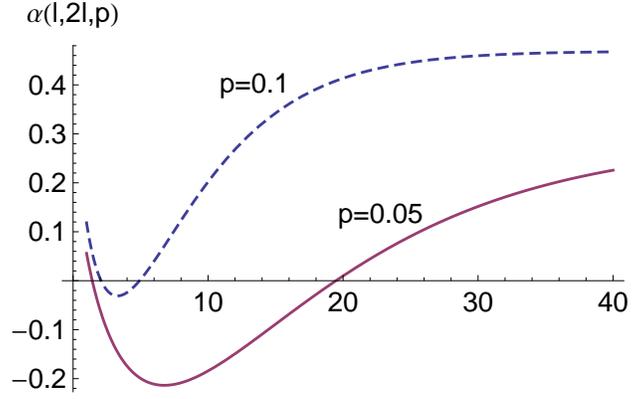}
\end{center}
\caption{Plot of $\alpha(l,r,p) =h(p)- \frac{l}{r}h((1-p)^r)$ as a function of $l$ $(r = 2l)$}
\label{funcr}
\end{figure}

We now discuss the estimation problem from an information-theoretic point of view.
This means that we allow the number of objects to increase up to infinity (i.e., $n \rightarrow \infty$)
and that we are interested in the existence of a sequence of pairs of a graph and an estimator that can achieve an arbitrarily small error probability.
We expect that placing the problem in an asymptotic setting will clarify the essence of the problem and shed new light on the 
behavior of a finite system.
From (\ref{asymbound}), it can be seen that $h(p) - (l/r) h((1-p)^r) \le 0$ should be satisfied 
in order to achieve an arbitrarily small error probability as $n \rightarrow \infty$.
It is natural to study the behavior of the function $\alpha(l,r,p)$  defined by
\begin{equation}
\alpha(l, r,p) \defeq h(p)- \frac{l}{r}h((1-p)^r).
\end{equation}
Figure \ref{funcr} shows the value of $\alpha(l,r,p)$ as a function of $l$.
The ratio $m/n=l/r$ is kept equal to $1/2$. The two curves in the figure correspond to the cases 
where $p=0.1$ and $p=0.05$.
It should be noted that $\alpha(l,r,p)$ takes negative values in a finite range around the 
minimum of $\alpha(l,r,p)$. Furthermore, from the plots in Fig.\ \ref{funcr}, it can be observed that  an arbitrarily small estimation error probability
for the noiseless system requires
a sparse pooling graph (i.e., pooling matrix).

\begin{figure}[htbp]
\begin{center}
\includegraphics[scale=1.0]{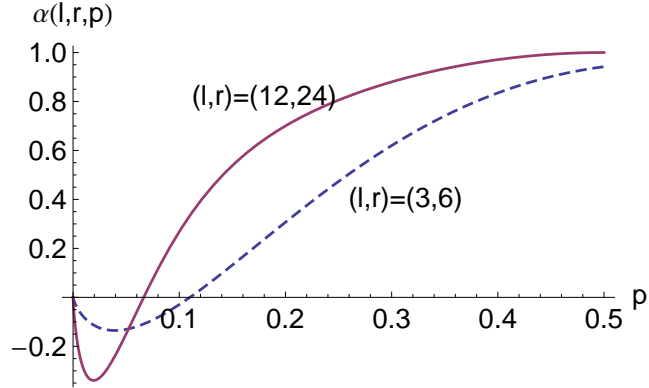}
\end{center}
\caption{Plot of $\alpha(l,r,p)=h(p)- \frac{l}{r}h((1-p)^r)$ as a function of $p$}
\label{funcp}
\end{figure}

Figure \ref{funcp} shows the value of $\alpha(l,r,p)$ as a function of $p$ for the two cases 
$(l,r)=(3,6)$ and $(l,r)=(12,24)$. In the case of the $(3,6)$-pooling graph,
$\alpha(3,6,p)$ takes a positive value if $p > p^*$, where $p^*$ is the positive root of 
$\alpha(3,6,p) = 0$ $(p^* \simeq 0.110023)$. This implies that in this case ($p > p^*$), an estimation with an arbitrarily small error probability $(n \rightarrow \infty)$
is impossible. Both of the curves have the same ratio $m/n = l/r = 1/2$.
The pooling graph with $(l,r)=(12,24)$ is worse than with $(l,r) = (3,6)$ in the sense that the
$(12,24)$ graph has a wider impossibility region.
This example indicates that a careful choice of parameters $(l,r)$ is required in order to design an appropriate 
pooling graph.

\subsection{Lower bound for noisy system}

Let us recall the problem setup for the noisy system.
The random variable $Z \defeq (Z_1,\ldots, Z_m)$, representing a noisy observation, is defined by 
\begin{equation}
Z = Y+E = F_G(X) + E.
\end{equation}
As in the case of the noiseless system, a lower bound on the error probability for the noisy system
can be derived based on Fano's inequality.
\begin{theorem}[Lower bound on estimation probability: noisy system]
Assume a noisy system.
For any pair of an $(l,r,n)$-pooling graph and an estimator,  
the error probability $P_e$ is bounded from below by
\begin{equation} \label{qineq}
h(p)+\frac l r  h(q)  - \frac{l}{r} h((1-p)^r (1-q) +(1-(1-p)^r) q) - \frac 1 n  \le P_e.
\end{equation}
\end{theorem}
(Proof) 
Based on the same argument as in the proof of Theorem \ref{th-noiseless}, we immediately have the inequality 
\begin{eqnarray}
n h(p) &\le& I(X;Z) + 1 + P_e n \\ \label{zineq}
&=& H(Z)- H(Z|X) + 1 + P_e n.
\end{eqnarray}

Since the variables $X,Y,Z$ constitute a Markov chain $X\rightarrow Y \rightarrow Z$, 
the data processing inequality 
$
I(X;Z) \le I(Y;Z)
$
holds, and it implies $H(Z|X) \ge H(Z|Y)$.
Applying $H(Z|X) \ge H(Z|Y)$ to (\ref{zineq}),  we can further rewrite the right-hand side of (\ref{zineq})
as follows: 
\begin{eqnarray}
n h(p) &\le& H(Z)- H(Z|Y) + 1 + P_e n \\
&=&H(Z)- m h(q) + 1 + P_e n \\ \label{dddf}
&\le&m H(Z_1)- m h(q) + 1 + P_e n.
\end{eqnarray}
The second line is due to the fact that $H(Z|Y) = m h(q)$, and the third line is based on the 
same argument that was used for deriving (\ref{Ybound}).
From the definition, the random variable $Z_1$ is given by
\begin{equation}
Z_1 = OR(X_i)_{i \in M(1)} + E_1.
\end{equation}
Since the random variable $E_1$ is a Bernoulli random variable with probability $q$,
we have
\begin{equation}
Pr(Z_1 = 0) = (1-p)^r (1-q) + (1-(1-p)^r) q
\end{equation}
and thus obtain 
\begin{equation}
H(Z_1) = h((1-p)^r (1-q) + (1-(1-p)^r) q).
\end{equation}
Substituting this into (\ref{dddf}), we obtain the following inequality 
\begin{equation} \label{qineq}
n h(p) \le m h((1-p)^r (1-q) + (1-(1-p)^r) q)- m h(q) + 1 + P_e n.
\end{equation}
The claim of the theorem is immediately derived from this inequality.
\hfill\qed

\begin{figure}[htbp]
\begin{center}
\includegraphics[scale=1.0]{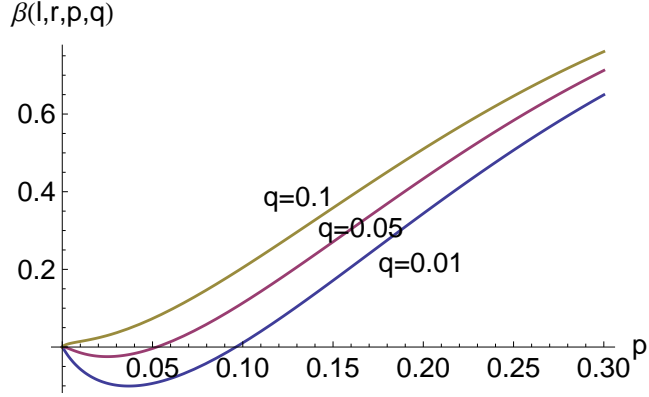}
\end{center}
\caption{Plot of $\beta(l,r,p,q) = h(p)+\frac l r  h(q)  - \frac{l}{r} h((1-p)^r (1-q) +(1-(1-p)^r) q) $ as a function of $p$ $(l=3, r=6)$}
\label{funcbeta}
\end{figure}

Note that by setting $q=0$, the lower bound (\ref{qineq}) is reduced to the lower bound (\ref{asymbound}) for 
the noiseless system.
In order to see the asymptotic behavior of the lower bound (\ref{qineq}), we plot $\beta(l,r,p,q)$ defined by
\begin{equation}
\beta(l,r,p,q) \defeq h(p)+\frac l r  h(q)  - \frac{l}{r} h((1-p)^r (1-q) +(1-(1-p)^r) q) 
\end{equation}
as a function of $p$, in Fig.\ \ref{funcbeta}.
It can be observed that, as $q$ increases, the value of $\beta(l,r,p,q)$ becomes larger.

\section{Direct Part Analysis}
\label{directpart}

In the previous section, we discussed the limitations of accurate estimation by any estimator,
i.e., a lower bound on the error probability. This result is similar to the converse part of a coding theorem.
In this section, we shall discuss the direct part, i.e., the existence of a sequence of estimators that can achieve an 
arbitrarily small error probability. As in the case of coding theorems, we here rely on the standard 
{\em bin-coding} argument \cite{Cover} to prove the main theorems. In order to apply such an information-theoretic 
argument, we will introduce a novel class of estimators, the {\em typical set estimators}.

\subsection{Pooling graph ensemble}

In the following analysis, we will take the average of the error probability of the typical set estimator over 
an ensemble of pooling graphs. The pooling graph ensemble introduced below resembles the 
bipartite graph ensemble for regular LDPC codes.
The following definition gives the details of the pooling graph ensemble \cite{modern} \cite{LS02}.
\begin{definition}[Pooling graph ensemble]
Let $G_{l,r,n}$ be the set of all $(l,r,n)$-regular bipartite graphs with $n$ left and $m=(l/r)n$ right nodes.
The cardinality of $G_{l,r,n}$ is $(n l)!$. Assume that equal probability $P(G) = 1/(nl)!$ is assigned for each graph $G \in G_{l,r,n}$.
The probability space based on the pair $(G_{l,r,n}, P)$ is called 
the $(l,r,n)$-pooling graph ensemble. \hfill\qed
\end{definition}

In order to prove the direct theorems, we need to evaluate the expectation of the number of typical sequences $x$
satisfying $y=F_G(x)$ over the $(l,r,n)$-pooling graph ensemble. The next lemma plays a crucial role in deriving the 
main theorems. 
\begin{lemma} \label{lemmacoeff}
Assume that $s \in [0,m]$ and  $w \in [0,n]$ are given.
Let $y_s \in \{0,1\}^m$ be a binary $m$-tuple with weight $s$, and let
$x_w \in \{0,1\}^n$ be a binary $n$-tuple with weight $w$.
The probability of the event $y_s = F_G(x_w)$ is given by
\begin{equation}
{\sf E}\left[{\Bbb I}[y_s = F_G(x_w) ]\right] = \frac{1}{{nl  \choose w l} } {\sf Coeff}[ ((1+z)^r-1)^s, z^{lw} ],
\end{equation}
where ${\sf Coeff}[g(z),z^i]$ represents the coefficient of $z^i$ in the polynomial $g(z)$.
The function ${\Bbb I}[cond]$ is the indicator function, which takes the value 1 if $cond$ is true and 0 otherwise.
\end{lemma}
(Proof) We here assume the socket model for a bipartite graph ensemble.
The quantity ${\sf E}\left[{\Bbb I}[y_s = F_G(x_w) ]\right]$ can be rewritten as follows: 
\begin{eqnarray} \nonumber
{\sf E}\left[{\Bbb I}[y_s = F_G(x_w) ]\right] &=& \sum_{G \in G_{l,r,n}} P(G) {\Bbb I}[y_s = F_G(x_w)] \\  \nonumber
&=& \frac{1}{(n l)! } | \{G \in G_{l,r,n} \mid y_s = F_G(x_w) \} | \\  \nonumber
&=& \frac{1}{(n l)! } {\sf Coeff}[ ((1+z)^r-1)^s 1^{m-s}, z^{lw} ] (l w)! ((n-w)l)! \\ \label{enumu}
&=& \frac{1}{{nl  \choose w l} } {\sf Coeff}[ ((1+z)^r-1)^s, z^{lw} ].
\end{eqnarray}
The first line is due to the definition of the expectation over the $(l,r,n)$-pooling graph ensemble.
Since $P(G)=1/(nl)!$ for any $G\in G_{l,r,n}$, we immediately have the second line.
The number of graphs satisfying $y_s = F_G(x_w)$ can be counted by using a generating function.
The set of right nodes with value $1$ is denoted by $R_1 (|R_1| = s)$, and the set of the remaining nodes is denoted by
$R_0 (|R_0|=m-s)$. The $i$-th coefficient of the product of 
generating functions $((1+z)^r-1)^s$ for $R_1$ and $1^{m-s}$ for $R_0$
represents the number of possible $(0,1)$-assignments with weight $i$ for the right sockets resulting $Y=y_s$.
There are $w$ left nodes with the value 1, which is assigned to the $l w$ left sockets.
Thus, the number of graphs satisfying $y_s = F_G(x_w)$ becomes the following product:
\begin{equation} \label{countingeq}
| \{G \in G_{l,r,n} \mid y_s = F_G(x_w) \} | =  {\sf Coeff}[ ((1+z)^r-1)^s, z^{lw} ] (l w)! ((n-w)l)!,
\end{equation} 
where  ${\sf Coeff}[ ((1+z)^r-1)^s, z^{lw} ]$ is the number of 
possible assignments of $(0,1)$ with weight $l w$ for the right sockets.
The number $(lw)!$ is the number of ways in which it is possible to connect the $l w$-left sockets (with the value 1) with the $l w$-right sockets (with the value 1).
The number $((n-w)l)!$ is the number of ways in which it is possible to connect the remaining left and right sockets.
The claim of this lemma is a consequence of the counting formula (\ref{countingeq}).
\hfill\qed

The combinatorial argument presented in the proof of  Lemma \ref{lemmacoeff} is closely related to the derivation of
an average input-output weight distribution of LDPC codes over a regular bipartite graph ensemble, presented by Hsu and  Anastasopoulos \cite{Hu}.

\subsection{Analysis on error probability for noiseless system}

In this subsection, we define the typical set estimator for the noiseless system and analyze its error performance.
Before describing the typical set estimator, we define the typical set \cite{Cover} as follows.
\begin{definition}[Typical set]
Assume that an i.i.d. random variables $A_i (i \in [1,n])$,
a positive constant $\epsilon$ and a positive integer $n$ are given.
The typical set $T_{n, \epsilon}$ is defined by
\begin{equation}
T_{n, \epsilon}  \defeq \left\{(a_1,\ldots, a_n) \in {\cal A}^n \mid 2^{-n({\cal H}+\epsilon)} \le Pr(a_1,\ldots, a_n) 
\le 2^{-n({\cal H}-\epsilon)} \right\},
\end{equation}
where ${\cal A}$ is the finite alphabet of $A_i$ and ${\cal H} \defeq H(A_i)$ holds for $i \in [1,n]$.
\hfill\qed
\end{definition}

The typical set estimator defined below is almost the same as the typical set decoder assumed in the proof of several coding theorems, such as in \cite{MacKay}.
It is exploited in order to simplify the proof, and it is, in general, computationally infeasible. Despite its computational complexity,
the performance of the typical set estimator can be used as a benchmark for other estimation algorithms.
In the following, we assume that ${\cal A} \defeq \{0,1\}$.

\begin{definition}[Typical set estimator]
Assume the noiseless system.
Suppose that an $(l,r,n)$-pooling graph $G \in G_{l,r,n}$ and a positive real value $\epsilon$ are given.
The typical set estimator $\Phi: \{0,1\}^m \rightarrow \{0,1\}^n \cup \{E\}$ is 
defined by
\begin{equation}
\Phi(y) \defeq
\left\{
\begin{array}{ll}
x \in D(y), & \mbox{if } |D(y)|=1, \\
E, & otherwise,
\end{array}
\right.
\end{equation}
where $D(y) (y \in \{0,1\}^m )$ is the decision set defined by
\begin{equation}
D(y) \defeq
\{x \in T_{n, \epsilon} \mid y = F_G(x)  \}.
\end{equation}
The symbol $E$ represents failure of the estimation.
\hfill\qed
\end{definition}
The typical set estimator $\Phi$ depends on the {\em bins} defined on the typical set $T_{n, \epsilon}$.
A bin $D(y)$ consists of the inverse image of $y$ in the typical set.
For an observed vector $y$, if the cardinality of the bin $D(y)$ is 1, the estimator declares that
$x \in D(y)$ has occurred. The estimation fails when the cardinality of $D(y)$ is greater than 1.
For evaluating the error probability of the typical set estimator, an analysis for this event is indispensable, and 
it will be the main topic of the following analysis.

The next lemma proves  the existence of a pair $(G, \Psi)$ achieving a given upper bound on 
the error probability. The proof of this lemma has a similar structure of 
the proof of the coding theorem for LDPC codes presented in \cite{MacKay}.
\begin{lemma} \label{noiselesslemma}
Assume the noiseless system.
If  $\gamma > 0$ satisfying 
\begin{equation} \label{noiselescond}
-(l-1) h(p)+
\max_{\sigma \in [0, l/r] } \left[
 \log_2 \inf_{z > 0}\frac{((1+z)^r-1)^\sigma}{z^{l p}} \right]+\gamma< 0
\end{equation}
exists, then there exists a pair $(G\in G_{l,n,r} , \Phi)$ for which
the error probability is smaller than $\gamma$.
\end{lemma}
(Proof) The proof is based on the bin-coding argument.
Assume that a positive real number $\epsilon$ is given (later we will see that $\epsilon$ is determined according to $\gamma$, but for now 
we consider that $\epsilon$ is given). 
Note that there are two events that the typical set estimator fails to correctly estimate.
By Event I, we denote the event in which a realization of $X$, $x$, is not a typical sequence.
Event II corresponds to the case in which a realization $x$ is a typical sequence, but $|D(F_G(x))| > 1$ holds.

We therefore have
\begin{equation}
P_e = Pr[X \ne \Phi (F_G(X))] = P_I + P_{II}(G),
\end{equation}
where $P_I$ and $P_{II}(G)$ are the probabilities corresponding to Events I and II, respectively.
Note that the probability $P_I$ depends only on the parameters $n$ and $\epsilon$.

We first consider the probability $P_{II}(G)$, for which the upper bound is as follows:
\begin{eqnarray}
P_{II}(G) 
&=& \sum_{x \in T_{n, \epsilon}} Pr(x) \Bbb I[ \exists x' \in T_{n,\epsilon}, x' \in D(F_G(x)), x' \ne x ] \\ \label{fgub}
&\le& \sum_{x \in T_{n, \epsilon}} Pr(x) \sum_{x' \in T_{n, \epsilon},  x \ne x'} \Bbb I[ F_G(x) = F_G(x') ].
\end{eqnarray}

By taking the expectation of (\ref{fgub}) over the $(l,r,n)$-pooling graph ensemble, we obtain
\begin{eqnarray} \label{fgfg}
{\sf E}[P_{II}(G) ]
&\le& \sum_{x \in T_{n, \epsilon}} Pr(x) \sum_{x' \in T_{n, \epsilon},  x \ne x'} {\sf E}[ \Bbb I[F_G(x) = F_G(x')] ] \\ \label{ysfg}
&\le&  |T_{n, \epsilon}|  \max_{s \in [0,m]}\max_{w \in [w_{min}, w_{max}]}  {\sf E}[ \Bbb I[ y_s = F_G(x_w)] ],
\end{eqnarray}
where $w_{min}$ and $w_{max}$ are defined by
\begin{equation}
w_{max} = \max_{x \in T_{n, \epsilon}}  wt(x), \quad w_{min} = \min_{x \in T_{n, \epsilon}}  wt(x),
\end{equation}
where $wt(x)$ represents the Hamming weight of $x$.
The vector $y_s$ is an arbitrary binary $m$-tuple with weight $s$, and $x_w$ is an arbitrary binary $n$-tuple 
with weight $w$.
The first inequality (\ref{fgfg}) is due to the linearity of the expectation.
In the derivation of (\ref{ysfg}), we used the inequality 
$\sum_{x \in T_{n, \epsilon}} Pr(x) \le 1$ .

Applying  the upper bound for the size of the typical set and Lemma \ref{lemmacoeff} to (\ref{ysfg}),  we have
\begin{equation} \label{pii}
{\sf E}[P_{II}(G) ]\le  2^{n(h(p) + \epsilon)}  \max_{s \in [0,m]}\max_{w \in [w_{min}, w_{max}]}  \frac{1}{{nl  \choose w l} } {\sf Coeff}[ ((1+z)^r-1)^s, z^{lw} ].
\end{equation}
By letting  $\omega \defeq w/n$ and $\sigma \defeq s/n$,  the above inequality (\ref{pii}) can be rewritten as
$
{\sf E}[P_{II}(G) ]\le  2^{n(h(p) + \epsilon + Q ) }
$,
where
\begin{eqnarray}\label{deltaomega}
Q &\defeq& \frac 1 n \log_2 \left[
 \max_{s \in [0,m]}\max_{w \in [w_{min}, w_{max}]}  \frac{1}{{nl  \choose w l} } {\sf Coeff}[ ((1+z)^r-1)^s, z^{lw} ]
 \right]  \\ 
&=&
\max_{\sigma \in [0, l/r] }\max_{\omega \in c(p,\epsilon) } \left[- h(\omega) 
+ \log_2 \inf_{z > 0}\frac{((1+z)^r-1)^\sigma}{z^{l \omega}} \right] + \delta(n).
\end{eqnarray}
For evaluating the coefficient of the generating function in (\ref{deltaomega}), a theorem by Burshtein and Miller \cite{Burshtein} is exploited.
Note that $\delta(n) \rightarrow 0$ as $n \rightarrow \infty$. The domain of 
$\omega$, $c(p, \epsilon)$,  is defined as
\begin{equation}
c(p, \epsilon) \defeq \{\omega \mid -h(p)-\epsilon \le \omega \log_2 p + (1-\omega) \log_2(1-p) \le - h(p) + \epsilon \}.
\end{equation}
It is clear that $\omega$ converges to $p$ if $\epsilon \rightarrow 0$, according to the domain $c(p, \epsilon)$.
This implies that $Q$ can be expressed as
\begin{eqnarray}
Q &=& - h(p)+
\max_{\sigma \in [0, l/r] }  \left[ 
 \log_2 \inf_{z > 0}\frac{((1+z)^r-1)^\sigma}{z^{l p}} \right] + \delta(n) + \xi(\epsilon),
\end{eqnarray}
where $\xi(\epsilon)$ is a function of $\epsilon$ such that $\xi(\epsilon) \rightarrow 0$ when $\epsilon \rightarrow 0$.
Assume that a positive real number $\gamma$ is given and 
\begin{equation}\label{negativecond}
-(l-1) h(p) + \max_{\sigma \in [0, l/r] }  \left[  \log_2 \inf_{z > 0}\frac{((1+z)^r-1)^\sigma}{z^{l p}} \right]  + \gamma < 0
\end{equation}
holds. 
For sufficiently large $n$ and sufficiently small $\epsilon$, there exists a pair $(n, \epsilon)$ satisfying 
$
\epsilon + \delta(n) + \xi(\epsilon) < \gamma
$
and the following two conditions. The first condition is that
\begin{eqnarray} \label{boundd}
{\sf E}[P_{II}(G) ] &\le&  2^{n \left( -(l-1) h(p) + \max_{\sigma \in [0, l/r] }
  \left[  \log_2 \inf_{z > 0}\frac{((1+z)^r-1)^\sigma}{z^{l p}} \right]  + \gamma \right) } \\
  &<& \frac{\gamma}{2}.
\end{eqnarray}
Note that, due to the assumption (\ref{negativecond}),  the exponential growth rate of the right-hand side of (\ref{boundd})
 is negative, and thus
the upper bound on ${\sf E}[P_{II}(G) ]$ can be arbitrarily small. The second condition 
is that 
$
P_I <  \gamma/2,
$
which is guaranteed by the asymptotic equipartition property (AEP) for the typical set \cite{Cover}. 
As a result, we have ${\sf E}[P_E] = P_I + {\sf E}[P_{II}(G) ] < \gamma$, and this implies the existence of 
a pair $(G\in G_{l,n,r} , \Phi)$ for which
the error probability is smaller than $\gamma$. \hfill\qed

In order to grasp the asymptotic behavior of the system, let us define $\theta_{l,r,p,\sigma}(z)$ by
\begin{equation}
\theta_{l,r,p,\sigma}(z) \defeq -(l-1) h(p)+ \log_2 \frac{((1+z)^r-1)^\sigma}{z^{l p}}.
\end{equation}
The condition in (\ref{noiselescond}) can be transformed  as 
\begin{eqnarray} 
-(l-1) h(p)+
\max_{\sigma \in [0, l/r] } \left[
 \log_2 \inf_{z > 0}\frac{((1+z)^r-1)^\sigma}{z^{l p}} \right]
 = 
\max_{\sigma \in [0, l/r] }\inf_{z > 0} \left[\theta_{l,r,p,\sigma}(z)  \right].
\end{eqnarray}

\begin{figure}[htbp]
\begin{center}
\includegraphics[scale=1.0]{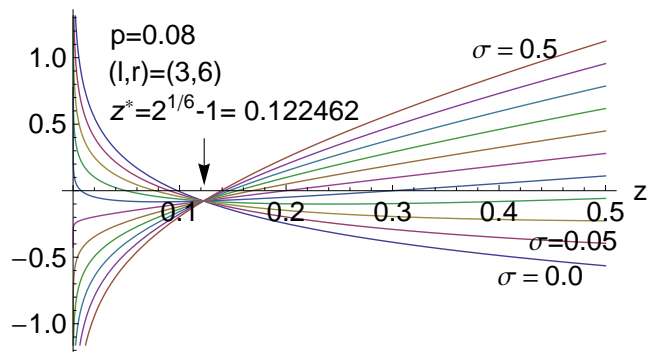}
\end{center}
\caption{Values of $\theta_{l,r,p,\sigma}(z) = -(l-1) h(p)+ \log_2 \frac{((1+z)^r-1)^\sigma}{z^{l p}}(l=3, r=6, p=0.08)$ as a function of $z$}
\label{direct1}
\end{figure}

Figure \ref{direct1} shows the plot of $\theta_{3,6,0.08, \sigma}(z)$ for $\sigma = 0.0, 0.05, 0.10,\ldots, 0.5$.
In this case, it is clear that $\max_{\sigma \in [0, l/r] }\inf_{z > 0} \left[\theta_{l,r,p,\sigma}(z)  \right] < 0$ holds.
We can observe that these curves intersect at a single point. At the intersection point, the value of $\theta_{3,6,0.08, \sigma}(z)$
is independent of the choice of $\sigma$ and thus $(1+z)^r-1$ should equal 1.
This means the solution of $(1+z)^r-1 = 1$, which is given by $z^*=2^{1/r}-1$,  gives the fixed point of $\theta_{3,6,0.08, \sigma}(z)$ 
in terms of $\sigma$.

This property of the intersection point is utilized in the following theorem to 
simplify the condition.
\begin{theorem}[Achievability of accurate estimation: noiseless system ] \label{noiselesstheorem}
Assume the noiseless system.
If $\gamma > 0$ satisfying 
\begin{equation}
-(l-1)h(p) - l p \log_2(2^{1/r} - 1) + \gamma < 0
\end{equation}
exists, then there exists a pair $(G\in G_{l,n,r} , \Phi)$ for which
the error probability is smaller than $\gamma$.
\end{theorem}
(Proof) 
Let $z^* \defeq 2^{1/r} - 1$.  
The condition (\ref{negativecond}) can be rewritten as
\begin{equation}
R  \defeq \max_{\sigma \in [0, l/r] } \inf_{z > 0} \left[-(l-1) h(p) +   \log_2 \frac{((1+z)^r-1)^\sigma}{z^{l p}} \right]  + \gamma < 0.
\end{equation}
Substituting $z^*$ into the right-hand side of $R$, we obtain an upper bound on $R$ as follows: 
\begin{eqnarray}
R &\le& \max_{\sigma \in [0, l/r] }\left[-(l-1) h(p) +   \log_2 \frac{((1+z^*)^r-1)^\sigma}{z^{* l p}} \right]  + \gamma \\
&=& \max_{\sigma \in [0, l/r] }\left[-(l-1) h(p)  + \log_2 \frac{ 1^\sigma}{(2^{1/r} - 1)^{l p}}  \right]  + \gamma \\
&=& \max_{\sigma \in [0, l/r] }\left[-(l-1) h(p)  - lp \log_2 (2^{1/r} -1) \right]  + \gamma \\
&=& -(l-1) h(p)  - lp \log_2 (2^{1/r} -1)  + \gamma. 
\end{eqnarray}
Thus, the condition $-(l-1) h(p)  - lp \log_2 (2^{1/r} -1)  + \gamma < 0$ implies $R < 0$,
and Lemma \ref{noiselesslemma} can be applied. \hfill\qed

\begin{figure}[htbp]
\begin{center}
\includegraphics[scale=1.0]{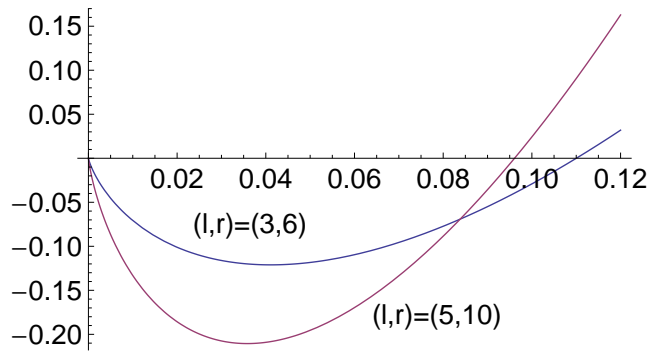}
\end{center}
\caption{Values of $\lambda(p)  = -(l-1)h(p) - l p \log_2(2^{1/r} - 1)$ as a function of $p$}
\label{direct2}
\end{figure}

Let 
\begin{equation}
\lambda(p) \defeq -(l-1)h(p) - l p \log_2(2^{1/r} - 1).
\end{equation}
The function $\lambda(p)$ is a convex function of $p$.
The equation $\lambda(p) = 0$ has only one solution $p^*$ in the range $0 < p^* < 1$ if $l > 1$ and $r > 1$.
Note that $\lambda(p) < 0$ if $0< p < p^*$.
Figure \ref{direct2} indicates the curves of $\lambda(p)$ for the cases $(l,r)=(3,6)$ and $(l,r) = (5,10)$.

\subsection{Threshold bounds}

From Theorem \ref{th-noiseless} (lower bound on the error probability) and Theorem \ref{noiselesstheorem},
it is natural to conjecture the existence of a threshold value $p^*(l,r)$ that partitions the range of $p$ into two regions.
Namely, if $p < p^*(l,r)$, an arbitrarily accurate estimation is possible. Otherwise, i.e., $p > p^*(l,r)$,
no estimator that can achieve an arbitrarily small error probability exists in the asymptotic limit $n \rightarrow \infty$.

An upper bound on the threshold  can be obtained from Theorem \ref{th-noiseless}.
The upper bound $p^*_U(l,r)$ is given by
\begin{equation}
p^*_U(l,r) \defeq \inf \left\{p  \mid p  \mbox{ satisfies } h(p) - (l/r) h((1-p)^r > 0 \right\}.
\end{equation}
On the other hand, a lower bound on the threshold is defined by
\begin{equation}
p^*_L(l,r) \defeq 
\sup \left\{p \mid p  \mbox{ satisfies} -(l-1)h(p) - l p \log_2(2^{1/r} - 1) < 0
 \right\},
\end{equation}
which is a direct consequence of Theorem \ref{noiselesstheorem}.
Table \ref{thresholdtable} presents the values of the lower and upper bounds on 
the threshold for the two cases $l/r=1/2$ and $l/r=1/4$.
From Table \ref{thresholdtable}, we can see that small gap between the lower and upper bounds still 
exists. However, the gap becomes fairly small when a pair $(l,r)$ gives the the largest value of $p^*_L(l,r)$.
For example, in the case of $l/r=1/2$, it can be observed that the pair $(3,6)$ gives the largest value.
In this case, the bounds are $P_L^*(3,6) =0.110022$ and $P_U^*(3,6) = 0.110023$ and the gap is 
approximately $10^{-6}$.

\begin{table}[htdp]
\caption{Threshold bounds for noiseless systems}
\label{thresholdtable}
\begin{center}
\begin{tabular}{c|ll}
\hline
$(l,r)$ & $P_L^*(l,r)$ &  $P_U^*(l,r)$ \\
\hline
$(2,4)$ & 0.092763 & 0.097350 \\
$(3,6)$ & 0.110022 & 0.110023 \\
$(4,8)$ & 0.104629 & 0.105999\\
$(5,10)$ & 0.096091 & 0.099480\\
$(6,12)$ & 0.087848 & 0.093027\\
\hline
$(2,8)$ & 0.022022 & 0.026824 \\
$(3,12)$ & 0.038651 & 0.039535 \\
$(4,16)$ & 0.041685 & 0.041687 \\
$(5,20)$ & 0.040693 & 0.040978 \\
$(6,24)$ & 0.038556 & 0.039427 \\
\hline
\end{tabular}
\end{center}

\end{table}%

\subsection{Analysis on error probability for noisy system}

In the previous subsection, we determined the accuracy of the best estimation that can be achieved for the noiseless system.
The argument used in the derivation of Theorem \ref{noiselesstheorem} can be extended to an argument for 
the noisy system. In this subsection, we will present the 
counterpart of Theorem \ref{noiselesstheorem}, an achievability theorem for the noisy system.

In the case of the noisy system, we again uses a typical set estimator but it needs to incorporate the effect of noises.
\begin{definition}[Typical set estimator]
Assume that an $(l,r,n)$-pooling graph $G \in G_{l,r,n}$ and positive real values $\epsilon_1$ and $\epsilon_2$ are given.
The typical set estimator $\Phi: \{0,1\}^m \rightarrow \{0,1\}^n \cup \{E\}$ is 
defined by
\begin{equation}
\Phi(y) \defeq
\left\{
\begin{array}{ll}
x \in D(y), & \mbox{if } |D(y)|=1, \\
E, & otherwise,
\end{array}
\right.
\end{equation}
where $D(y) (y \in \{0,1\}^m )$ is the decision set defined by
\begin{equation}
D(y) \defeq
\{x \in T_{n, \epsilon_1} \mid y = F_G(x) + e, e \in T_{m,\epsilon_2}  \}.
\end{equation}
\hfill\qed
\end{definition}
It should be remarked that, in this case,  a bin $D(y)$ is defined by the set of 
$x \in T_{n, \epsilon_1}$ satisfying $y = F_G(x) + e (e \in T_{m,\epsilon_2})$, which
includes the effect of an additional {\em typical noise} $e$.

Then next lemma is required to prove the achievability theorem, which will be presented below.
The proof of the lemma is based on a generating function method.

\begin{lemma} \label{noisylemma}
Assume the noisy system, and assume that $s \in [0,m]$ and  $w \in [0,n]$ are given.
Let $y_s \in \{0,1\}^m$ be a binary $m$-tuple with weight $s$, and let
$x_w \in \{0,1\}^n$ be a binary $n$-tuple with weight $w$.
The following equality 
\begin{equation} \label{zetaeq}
\sum_{e \in \{0,1\}^m } Pr(e) {\sf E}\left[{\Bbb I}[y_s = F_G(x_w) \oplus e]\right] 
= \frac{1}{{nl  \choose w l} } {\sf Coeff}[\zeta(z) , z^{lw} ]
\end{equation}
holds where
\begin{equation}
\zeta(z) \defeq (((1+z)^r-1)(1-q)+q)^s (((1+z)^r-1)q+(1-q))^{m-s}
\end{equation}
and $Pr(e) = q^{wt(e)}(1-q)^{m -wt(e)}$. The expectation in (\ref{zetaeq}) is taken over the $(l,r,n)$-pooling graph ensemble.
The operator $\oplus$ represents the mod-2 addition (exclusive OR) operator.
\end{lemma}
(Proof) 
The left-hand side of (\ref{zetaeq}) can be rewritten as follows.
\begin{eqnarray} \nonumber
&&\hspace{-2.5cm}\sum_{e \in \{0,1\}^m } Pr(e) {\sf E}\left[{\Bbb I}[y_s = F_G(x_w) \oplus e]\right] \\ \label{pre}
&=& \sum_{e \in \{0,1\}^m } Pr(e)\sum_{G\in G_{l,r,n}} Pr(G) {\Bbb I}[y_s = F_G(x_w) \oplus e] \\ \label{prepre}
&=& \sum_{e \in \{0,1\}^m }Pr(e)\frac{1}{(n l)! } \# \{G \in G_{l,r,n} \mid y_s = F_G(x_w) \oplus e \} \\ \label{preprepre}
&=& \frac{1}{(n l)! } {\sf Coeff}[ \zeta(z), z^{lw} ] (l w)! ((n-w)l)! \\
&=& \frac{1}{{nl  \choose w l} } {\sf Coeff}[ \zeta(z), z^{lw} ].
\end{eqnarray}
The equalities (\ref{pre}) and (\ref{prepre}) are due to the definitions of the expectation and the ensemble.
The derivation of (\ref{preprepre}) is based on the following argument. 
A right node having the value of 1 corresponds to the generating function
$
((1+z)^r-1)(1-q)+q,
$
which is the weighted sum of two generating functions for the cases $(y_i = 1, e_i =0)$ (i.e., $(1+z)^r-1$ with probability $1-q$)
and $(y_i = 0, e_i =1)$ (i.e., $1$ with probability $q$).
Similarly, a right node having the value of 0 corresponds to the generating function
$
((1+z)^r-1)q+(1-q),
$
which is the weighted sum of two generating functions for the cases $(y_i = 0, e_i=0)$ and $(y_i = 1, e_i=1)$.
We again assume the socket model. Applying almost the same argument as was used in the proof of Lemma \ref{lemmacoeff}, 
we have 
\begin{equation}
\sum_{e \in \{0,1\}^m }Pr(e)  \# \{G \in G_{l,r,n} \mid y_s = F_G(x_w) \oplus e \} 
=  {\sf Coeff}[ \zeta(z), z^{lw} ] (l w)! ((n-w)l)!.
\end{equation}
Due to this equation, we obtain (\ref{preprepre}). \hfill\qed

The following lemma provides for the achievability of an accurate estimation based on the typical set estimator.
\begin{lemma} \label{noisylemma}
Assume the noisy system.
If $\gamma > 0$ satisfying 
\begin{equation}
 -(l-1) h(p)+ \frac{l}{r} h(q) + \max_{\sigma \in [0, l/r]}
  \left[\log_2 \inf_{z > 0} \frac{\eta(z)}{z^{lp}}   \right] +\gamma < 0
\end{equation}
exists, then there exists a pair $(G\in G_{l,n,r} , \Phi)$ with 
the error probability smaller than $\gamma$.
The function $\eta(z)$ is defined by 
\begin{equation}
\eta(z) \defeq (((1+z)^r-1)(1-q)+q)^\sigma (((1+z)^r-1)q+1-q)^{1-\sigma}.
\end{equation}
\end{lemma}

(Proof)
There are three events in which the typical set estimator fails to output the correct estimate, as follows.

Event I occurs when $x$ is not a typical sequence (i.e., x $\notin T_{n, \epsilon_1}$).
The probability of Event I is denoted by $P_I$, which does not depend on $G$ and, due to the AEP, 
can be arbitrarily small as $n$ goes to infinity. 
Event II occurs when $x \in T_{n, \epsilon_1}$ holds but
$e \notin T_{m,\epsilon_2}$. Namely, in this case, the noise $e$ is not a typical sequence.
 In such a case, the estimator fails to give a correct estimate.
The probability for this event, $P_{II}$, can also be made arbitrarily small as $m$ goes to infinity, due to the AEP.
The last event, Event III, occurs when $x \in T_{n, \epsilon_1}$ and  $e \in T_{m, \epsilon_2}$, but
the cardinality of $D(F_G(x)+e)$ is greater than 1.
We can obtain an upper bound for the probability of Event III, $P_{III}(G)$, in the following way:
\begin{eqnarray} \nonumber
P_{III}(G)  &=&\sum_{x \in T_{n, \epsilon_1}} 
Pr(x) \sum_{e \in T_{m, \epsilon_2}} Pr(e)\Bbb I[ \exists x' \in T_{n,\epsilon}, x' \in D(F_G(x) \oplus e), x' \ne x ] \\ \nonumber
&\le& \sum_{x \in T_{n, \epsilon_1}} Pr(x)  \sum_{e \in T_{m, \epsilon_2}} Pr(e) 
\sum_{\stackrel{x' \in T_{n, \epsilon_1}}{ x \ne x'}}\sum_{e' \in T_{m, \epsilon_2}} \Bbb I[ F_G(x) \oplus e = F_G(x') \oplus e'] \\ \nonumber
&=&\sum_{x \in T_{n, \epsilon_1}} Pr(x)  \sum_{e \in T_{m, \epsilon_2}} Pr(e)  
 \sum_{\stackrel{x' \in T_{n, \epsilon_1}}{x \ne x'}}2^{m(h(q)+\epsilon_2 )}\left(\sum_{e' \in T_{m, \epsilon_2}} 2^{-m(h(q)+\epsilon_2 )}
 \Bbb I[ F_w(x) \oplus e = F_w(x') \oplus e'] \right) \\  \label{noisylast}
&\le& 2^{m(h(q)+\epsilon_2 )}\sum_{x \in T_{n, \epsilon_1}} Pr(x)  \sum_{e \in T_{m, \epsilon_2}} Pr(e) 
\sum_{\stackrel{x' \in T_{n, \epsilon_1}}{x \ne x'}} 
\left(\sum_{e' \in \{0,1\}^m} Pr(e')\Bbb I[F_G(x) \oplus e = F_G(x') \oplus e'] \right). 
\end{eqnarray}
The last inequality (\ref{noisylast}) is based on the inequalities
\begin{eqnarray} \nonumber
\sum_{e' \in T_{m, \epsilon_2}} 2^{-m(h(q)+\epsilon_2 )} t(e') &\le &
\sum_{e' \in T_{m, \epsilon_2}} Pr(e')t(e') \\
&\le& \sum_{e' \in \{0,1\}^m} Pr(e')t(e'),
\end{eqnarray}
where $t(e')=\Bbb I[F_G(x) \oplus e = F_G(x') \oplus e']$. 
By taking the expectation of $P_{III}(G)$ over  the $(l,r,n)$-pooling graph ensemble, 
we have 
\begin{eqnarray} \nonumber
{\sf E}[P_{III}(G) ] &\le&
2^{n(h(p)+\epsilon_1 )} 2^{m(h(q)+\epsilon_2 )}  
\max_{s \in [0,m]}\max_{w \in [w_{min}, w_{max}]} \left(\sum_{e' \in \{0,1\}^m} Pr(e')
{\sf E}[\Bbb I[ y_s = F_G(x_w) \oplus e']] \right) \\ \nonumber
&=&2^{n(h(p)+\epsilon_1 + \frac m n h(q) + \frac m n \epsilon_2  )} 
\max_{s \in [0,m]}\max_{w \in [w_{min}, w_{max}]}\frac{1}{{nl  \choose w l} } {\sf Coeff}[ \zeta(z), z^{lw} ].
\end{eqnarray}
The last equality is due to Lemma \ref{noisylemma}. The remaining argument is almost the same as was used in the proof of 
Lemma \ref{noiselesslemma}. \hfill\qed

The simplification presented in the proof of Theorem \ref{noiselesstheorem}
can also be applied to Lemma \ref{noisylemma}. The next theorem is the main contribution of this section.
\begin{theorem}[Achievability of accurate estimation: noisy system ] \label{noisytheorem}
Assume the noisy system.
If $\gamma > 0$ satisfying 
\begin{equation} \label{condnoisy}
-(l-1)h(p) + \frac{l}{r} h(q) -  l p \log_2(2^{1/r} - 1) + \gamma < 0
\end{equation}
exists, then there exists a pair $(G\in G_{l,n,r} , \Phi)$ for which
the error probability is smaller than $\gamma$.
\end{theorem}
(Proof) Let $z^* \defeq 2^{1/r} -1$.  It is easy to check that $\eta(z^*) = 1$ holds.
The argument used in the proof of Theorem \ref{noiselesstheorem} can be applied to Lemma \ref{noisylemma}. \hfill\qed

Comparing the conditions in Theorems \ref{noiselesstheorem} and \ref{noisytheorem},  we can see that
an additional term $(l/r) h(q)$ appears in (\ref{condnoisy}) in Theorem \ref{noisytheorem}. This additional 
term disappears if $q=0$, and in that case, the condition in Theorem \ref{noisytheorem} becomes identical to 
that in Theorem \ref{noiselesstheorem}.

\section{Generalization to sparse observation system}
\label{generalize}

In the previous sections, we have discussed the two systems, i.e.,
the noiseless and the noisy system. In the noiseless system, an observation (a test result)
consists of a $r$ binary values  that serve as input to
a test function (logical OR in the above cases) that is chosen 
according to the setting of the conventional group testing.
The argument for the theorems presented above can be naturally extended to 
those for more general classes of test functions. For example, 
it is desirable that a test function that outputs multiple values (such as 
negative, weak positive, positive, and strong positive) can be handled in a coherent manner.
In this section, we will discuss such an extension.

\subsection{Problem setup for generalized noiseless system}

In this subsection, we will explain the problem setup for the {\em generalized noiseless system}.

Let $X \defeq (X_1, X_2,\ldots, X_n)$ be the status vector for objects where  $X_i (i \in [1,n])$ is 
an i.i.d. $u$-ary random variable; i.e., 
$X_i$ takes a value in the alphabet ${\cal A} \defeq \{a_1,a_2,\ldots, a_u\} \subset \Bbb R$ with 
probability $Pr[X_i = a_k] = p_k (k \in [1,u])$. This means that $X$ can be considered as an output of 
a discrete memoryless source (DMS).
The test results are represented by the random variable $Y \defeq (Y_1,Y_2,\ldots, Y_m)$ 
where $Y_i (i \in [1,n])$ takes the value in the alphabet ${\cal B} \defeq \{b_1, b_2, \ldots, b_v\} \subset \Bbb R$. 
Suppose that a pooling graph $G \in G_{l,r,n}$ is given.

A test function $f: {\cal A}^r \rightarrow {\cal B}$ is assumed to be given as well.
We here assume that the test function $f$ is a symmetric function; i.e., 
$f(x) = f( \sigma(x) )$ holds for any $x \in {\cal A}^r$  and for any permutation $\sigma$ on $r$-arguments.
For $j \in [1,m]$, the test result $Y_j$ is given by
$
Y_j = f(X_i)_{i \in M(j)}.
$
Although it might be abuse of notation, we will write the functional relationship between $X$ and $Y$ by $Y=F_G(X)$
as well as the noiseless case discussed in the previous sections.

Let ${\cal S}\defeq \{s_1, s_2,\ldots, s_{|{\cal S}|}\} \subset \Bbb R$.
For a vector $x \in {\cal S}^\ell$, the number of appearances of $a \in {\cal S}$  in $x$ is denoted by $N_{\cal S}(a|x)$,
and the type vector of $x$ is defined by
$
N_{\cal S}(x) \defeq (N_{\cal S}(s_1|x), N_{\cal S}(s_2|x), \ldots, N_{\cal S}(s_{|{\cal X}|} |x)).
$
From this definition, it is evident that 
$
\sum_{a \in {\cal S}}  N_{\cal S}(a |x) = \ell 
$
holds.  
The type set $U_\ell^{({\cal S})}$ is given by
\begin{equation}
U_\ell^{({\cal S})} \defeq \left\{(t_1,t_2,\ldots, t_{|{\cal S}|}) \in [0,\ell]^{|{\cal S}|} \mid \sum_{i=1}^{|{\cal S}|} t_i = \ell   \right\}.
\end{equation}

The generating function defined below plays an important role in the following discussion.
\begin{definition}[generating functions for test function]
For $k \in [1,v]$, a generating function $\alpha_k(z_1, z_2, \ldots, z_u)$ is defined by
\begin{equation}
\alpha_k(z_1, z_2, \ldots, z_u) \defeq \sum_{x \in {\cal A}^r} \Bbb I[b_k = f(x)] z_1^{N_{\cal A}(a_1|x)} z_2^{N_{\cal A}(a_2|x)} \cdots z_u^{N_{\cal A}(a_u|x)}.
\end{equation}
\hfill\qed
\end{definition}

By using these generating functions, a lower bound for the estimation error probability 
can be compactly expressed as in the following theorem.
\begin{theorem}[Lower bound on estimation probability: generalized noiseless system] \label{th-gneralized}
Assume the generalized noiseless system.
For any pair of an $(l,r,n)$-pooling graph and an estimator  $(G,\Phi)$, 
the error probability $P_e$ is bounded from below by
\begin{equation}
h(p_1,\ldots, p_u) + \frac{l}{r} \sum_{k=1}^v \alpha_k(p_1, \ldots, p_u) \log_2 \alpha_k(p_1, \ldots, p_u) - \frac 1 n \le P_e,
\end{equation}
where $h(p_1,\ldots, p_u)$ is the entropy function defined by
\begin{equation}
h(p_1,\ldots, p_u) \defeq - \sum_{i=1}^u p_i \log_2 p_i.
\end{equation}
\end{theorem}
(Proof)
A realization vector of input $x \in {\cal A}^n$ is an output from the DMS with probability $Pr(X_i = a_j)=p_j (j \in [1,u])$.
Thus, the probability for $x$ is given by
\begin{equation}
Pr(x) = p_1^{N_{\cal A}(a_1|x)} p_2^{N_{\cal A}(a_2|x)} \cdots p_u^{N_{\cal A}(a_u|x)}.
\end{equation}
Due to the definition of the generator functions, we have 
\begin{eqnarray}
Pr(Y_i = b_k) &=& 
\sum_{x \in {\cal A}^r} \Bbb I[b_k = f(x)] p_1^{N_{\cal A}(a_1|x)} p_2^{N_{\cal A}(a_2|x)} \cdots p_u^{N_{\cal A}(a_u|x)} \\
&=& \alpha_k(p_1, p_2, \ldots, p_u).
\end{eqnarray}
The proof of this theorem is almost same as that of Theorem \ref{th-noiseless}. The difference 
is the evaluation of $H(Y_j) (j \in [1,m])$. 
For any $j \in [1,m]$, the entropy $H(Y_j)$ can be evaluated as
\begin{eqnarray} \nonumber
H(Y_j) 
&=& - \sum_{k=1}^{v} Pr (Y_j = b_k) \log_2 Pr (Y_j = b_k) \\ \nonumber
&=& - \sum_{k=1}^v \alpha_k(p_1, \ldots, p_u) \log_2 \alpha_k(p_1, \ldots, p_u)
\end{eqnarray}
The remaining discussion is almost same as that of the proof of Theorem \ref{th-noiseless}. \hfill\qed

The next lemma, which can be considered as a generalization of Lemma \ref{lemmacoeff}, 
provides the expectation of ${\Bbb I}[y_t = F_G(x_w) ]$ over the $(l,r,n)$-pooling graph ensemble.
The proof of this lemma is based on a combinatorial argument very similar to that of the binary group testing case.
\begin{lemma} \label{lemmacoeff2}
Assume that 
$
w = (w_1,w_2,\ldots, w_u) \in U_n^{({\cal A} )},
$
and
$
s=(s_1,s_2, \ldots, s_v)  \in U_m^{({\cal B} )}
$
are given.
Let $y_s \in {\cal B}^m$ be any $m$-tuple with type $s$ and 
$x_w \in {\cal A}^n$ be any $n$-tuple with type $w$.
The expectation of ${\Bbb I}[y_s = F_G(x_w) ]$ is given by
\begin{equation}
{\sf E}\left[{\Bbb I}[y_s = F_G(x_w) ]\right] = \frac{1}{{nl  \choose w_1 l \ w_2 l \cdots w_u l } } {\sf Coeff} \left[ \prod_{k=1}^v
\left(\alpha_k(z_1,\ldots, z_u) \right)^{s_i} , z_1^{l w_1}z_2^{l w_2} \cdots z_u^{l w_u}  \right].
\end{equation}
\hfill\qed 
\end{lemma}

The next theorem is a generalized version of the direct part theorem indicating a condition for
arbitrary accurate estimation.
\begin{theorem}[Achievability of accurate estimation: generalized system ] 
\label{gentheorem}
Assume the generalized noiseless system.
If  $\gamma > 0$ satisfying 
\begin{equation}\label{condgen}
-(l-1) h(p_1,\ldots, p_u) + \inf_{z_1,\ldots, z_u > 0} \left[ \frac l r  \max_{k=1}^v \left(\log_2 \alpha_k(z_1,\ldots, z_u)\right)    - \sum_{i=1}^u l p_i \log_2 z_i \right] + \gamma < 0,
\end{equation}
exists then there exists a pair $(G\in G_{l,n,r} , \Phi)$ with  the error probability smaller than $\gamma$.
\end{theorem}
(Proof)
The proof is almost same as that of Theorem \ref{noiselesstheorem}. We therefore focus on the 
difference. Due to the same argument discussed in the proof of Theorem \ref{noiselesstheorem}
and Lemma \ref{lemmacoeff2}, the expectation of the probability of Event II can be 
upper bounded by
\begin{eqnarray}
{\sf E}[P_{II}(G)]  \le 2^{n(h(p_1,\ldots, p_u)+\epsilon)} \max_{s \in T_m^{({\cal B})}} \max_{w \in \Delta} 
\frac{1}{{nl  \choose w_1 l \ w_2 l \cdots w_u l } } {\sf Coeff} \left[ \prod_{k=1}^v
\left(\alpha_k(z_1,\ldots, z_u) \right)^{s_i} , z_1^{l w_1}z_2^{l w_2} \cdots z_u^{l w_u}  \right].
\end{eqnarray}
The set $\Delta$ is defined by 
$
\Delta \defeq \{ N_{\cal A}(x) \in U_{n}^{{({\cal A})}} \mid x \in T_{n, \epsilon} \}
$
which is the set of possible type vectors in the typical set $T_{n, \epsilon}$.
Note that a type vector in $\Delta$ converges to $(p_1,\ldots, p_u)$
as $n \rightarrow \infty$ and $\epsilon \rightarrow 0$.
From this upper bound, the condition
\begin{equation} \label{naturaleq}
-(l-1)h(p_1,\ldots, p_u) + \max_{\sigma \in S(l,r,v)} \left[\log_2  \inf_{z_1,\ldots, z_u>0}
 \frac{\prod_{k=1}^v (\alpha_k(z_1,\ldots, z_u))^{\sigma_k}  }{z_1^{lp_1} \cdots z_u^{lp_u}  } \right] + \gamma < 0,
\end{equation}
is naturally derived by considering the exponential growth rate of the upper bound
where $\sigma\defeq(\sigma_1,\ldots, \sigma_v)$ and 
\begin{equation}
S(l,r,v) \defeq \left\{(s_1,s_2,\ldots, s_v) \in \Bbb R^v \mid s_k \ge 0 (k \in [1,v]), \sum_{k=1}^v s_k = l/r  \right\}.
\end{equation}
The exponential growth rate of the coefficient of the generator function
is derived based on a variation of the theorem by Burshtein and Miller \cite{Burshtein}.
The second term of (\ref{naturaleq}) can be upper bounded by
\begin{eqnarray} \nonumber
&&\hspace{-3cm}\max_{\sigma \in S(l,r,v)} \left[ \log_2  \inf_{z_1,\ldots, z_u>0}
 \frac{\prod_{k=1}^v (\alpha_k(z_1,\ldots, z_u))^{\sigma_k}  }{z_1^{lp_1} \cdots z_u^{lp_u}  } \right]  \\
&\le& 
  \inf_{z_1,\ldots, z_u>0}\left[
   \max_{\sigma \in S(l,r,v)} \left(\log_2 \frac{\prod_{k=1}^v (\alpha_k(z_1,\ldots, z_u))^{\sigma_k}  }{z_1^{lp_1} \cdots z_u^{lp_u}  } \right) \right] \\
&=&
  \inf_{z_1,\ldots, z_u>0} \left[\max_{\sigma \in S(l,r,v)}\left( \sum_{k=1}^v {\sigma_k} \log_2 \alpha_k(z_1,\ldots, z_u) \right) - l \sum_{i=1}^u p_i \log_2 z_i   \right].
\end{eqnarray}
Assume that $z_1,\ldots, z_u$ are fixed to certain real values.
In such a case, the maximum of 
\begin{equation}
\max_{\sigma \in S(l,r,v)}\left( \sum_{k=1}^v {\sigma_k} \log_2 \alpha_k(z_1,\ldots, z_u) \right)
\end{equation}
is obtained by setting
\begin{equation}
\sigma_{i} =
\left\{
\begin{array}{ll}
l/r,  & i = \arg \max_{k \in [1,v]} \left( \log_2 \alpha_k(z_1,\ldots, z_u) \right) \\
0, & \mbox{otherwise}.
\end{array}
\right.
\end{equation}
This gives the condition (\ref{condgen}) in the claim.
\hfill\qed

It might be useful to derive a corollary for 
the special case where inputs are binary random variables with $Pr(X_i = 0) = 1-p, Pr(X_i=1) = p$.
The condition in the corollary becomes much simpler than that of the previous theorem.
\begin{corollary}\label{coro}
Assume the generalized noiseless system with binary input alphabet ${\cal A} = \{a_1=0, a_2 = 1\}$.
If  $\gamma > 0$ satisfying 
\begin{equation}
-(l-1) h(p) + \inf_{z > 0} \left[ \frac l r  \max_{k=1}^v \left(\log_2 \beta_k(z)\right)   -  l p \log_2 z \right] + \gamma < 0,
\end{equation}
exists then there exists a pair $(G\in G_{l,n,r} , \Phi)$ with  the error probability smaller than $\gamma$. The 
generating functions $\beta_k(z)$ is defined by
\begin{equation}
\beta_k(z) \defeq \sum_{x \in \{0,1\}^r } \Bbb I[b_k = f(x)] z^{wt(x)}.
\end{equation}
\end{corollary}
(Proof)
Assume that $p_1 = 1-p$ and $p_2 = p$. The left hand side of (\ref{condgen}) is given by 
\begin{equation}\label{inside}
-(l-1) h(p) + \inf_{z_1,z_2 > 0} \left[ \frac l r  \max_{k=1}^v \left(\log_2 \alpha_k(z_1, z_2)\right)    - \sum_{i=1}^2 l p_i \log_2 z_i \right] + \gamma 
\end{equation}
in this case. From the definition of the generator functions $\alpha_k(z_1,\ldots, z_u)$, we have 
\begin{eqnarray}
\alpha_k(z_1,z_2) 
&=& \sum_{x \in \{0,1\}^r } \Bbb I[b_k = f(x)] z_1^{N_{\cal A}(0|x)} z_2^{N_{\cal A}(1|x)} \\
&=& \sum_{x \in \{0,1\}^r } \Bbb I[b_k = f(x)] z_1^{r - wt(x)} z_2^{wt(x)} \\ \label{wt}
&=& z_1^r \sum_{x \in \{0,1\}^r } \Bbb I[b_k = f(x)]  \left(\frac{z_2}{z_1}\right) ^{wt(x)}.
\end{eqnarray}
The objective function of infimum in (\ref{inside}) can be  simplified as follows:
\begin{eqnarray}
\frac l r  \max_{k=1}^v \left(\log_2 \alpha_k(z_1, z_2)\right)    - \sum_{i=1}^2 l p_i \log_2 z_i
&=& 
\frac l r  \max_{k=1}^v \left(\log_2 \alpha_k(z_1, z_2)\right)    -  l  \log_2 z_1 - p l \log_2 \left(\frac{z_2}{z_1} \right) \\
&=& 
\frac l r  \max_{k=1}^v \left(\log_2 \sum_{x \in \{0,1\}^r } \Bbb I[b_k = f(x)]  \left(\frac{z_2}{z_1}\right) ^{wt(x)}\right)   - p l \log_2 \left(\frac{z_2}{z_1} \right) \\
&=& 
\frac l r  \max_{k=1}^v \left(\log_2 \beta_k(z)  \right)   - p l \log_2 z.
\end{eqnarray}
In the derivation of the second equality, (\ref{wt}) was used.  
The replacement $z \defeq z_2/z_1$ and the definition of $\beta_k(z)$ yield the last equality.
\hfill \qed

As an application of this corollary,  we will derive the condition in 
Theorem \ref{noiselesstheorem} which states the achievability of accurate estimation 
for the noiseless system discussed in Section \ref{directpart}.
As stated before, the generator functions for the noiseless system
are given by $\beta_1(z)=1$ and $\beta_2(z)=(1+z)^r-1$. 
From Theorem \ref{gentheorem}, we thus have the achievability condition 
\begin{equation} \label{achicon}
-(l-1) h(p) + \inf_{z>0}  \left(\Bbb I[z \ge 2^{1/r} -1] \frac l r  \log_2((1+z)^r-1)     - lp \log_2 z \right) + \gamma < 0.
\end{equation}
The condition can be simplified into 
\begin{equation}
-(l-1) h(p)  - lp \log_2 (2^{1/r} -1) + \gamma < 0, 
\end{equation}
which is the achievability condition of Corollary \ref{coro}. It should be remarked that 
the function 
\[
\Bbb I[z \ge 2^{1/r} -1] (l/r)  \log_2((1+z)^r-1)  - lp \log_2 z
\]
appeared in (\ref{achicon}) 
corresponds to the upper envelope of the curves presented in Fig.\ref{direct1}.

\section{Conclusion}

There are strong similarities between 
group testing schemes and  linear error correction schemes for 
binary symmetric channels. The analysis presented in this paper was inspired by the 
theoretical works on LDPC codes \cite{MacKay} \cite{Burshtein2}.
From numerical evaluation, it was shown that 
the gap between the upper bound $p^*_U(l,r)$ and the lower bound $p^*_L(l,r)$  is usually quite small.
This suggests the existence of a sharp threshold, which is similar to the Shannon limit 
for a channel coding problem. 

On the other hand, from the analysis presented here, we note a fundamental difference between 
a group testing scheme and a binary linear error correction scheme.
The lower bounds on the estimation error probability, which we
proved in this paper, imply that
a group testing scheme requires a sparse pooling graph or pooling matrix 
for asymptotically optimal reconstruction. Note from an information-theoretic point of view, a dense parity check matrix is desirable for 
realizing a good linear error correction scheme.

The main contribution of this work is the direct theorems, which give the conditions 
for achieving arbitrarily small estimation error probabilities. In the same way that the Shannon limit has been used 
as a benchmark for the combination of a code and a decoding algorithm, these results can provide a 
concrete benchmark for existing and emerging reconstruction algorithms, such as belief propagation \cite{Sejdinovic} \cite{Kanamori},
and can stimulate the development of novel reconstruction algorithms.
In this paper, we assumed a Bernoulli source and noise, but, since the AEP holds for them, the arguments presented here may be 
applicable to stationary ergodic sources and noises.

\section*{Acknowledgement}
The author would like to thank the anonymous reviewers of International Symposium on Information Theory 2013
for their constructive comments on this work.
This work was supported by JSPS Grant-in-Aid for Scientific Research (B) Grant Number 25289114.

\end{document}